\documentclass[11pt]{article}

\newcommand{\remove}[1]{}
\newtheorem{theorem}{Theorem}
\newtheorem{lemma}{Lemma}
\newtheorem{cor}{Corollary}
\newtheorem{dfn}{Definition}

\newcommand{\zozo}{\{0,1\}}
\newcommand{\zo}[1]{\{0,1\}^{#1}}
\newcommand{\F}[1]{GF{#1}}

\newcommand{\Mn}{M_n(GF(2))}
\newcommand{\Mnp}{M_n(GF(p))}
\newcommand{\sq}{q_{*}(MP_n)}
\newcommand{\sbl}{{bl}_{*}(MP_n)}
\newcommand{\tbl}{{bl}_{tot}(MP_n)}

\def\QED{\hfill $\!\clubsuit$ \newline}
\def\cross{\!\times\!}

\newenvironment{proof}{\noindent{\bf Proof:}}{{\QED}}
\newenvironment{proofof}{\noindent{\bf Proof of}}{{\QED} }

\begin{document}

\title{Lower Bounds for Matrix Product}

\author{Amir Shpilka\\
Institute of Computer Science\\
Hebrew University\\ Jerusalem, Israel\\ amirs@cs.huji.ac.il \\}

\maketitle
\thispagestyle{empty}

\begin{abstract}
We prove lower bounds on the number of product gates in bilinear and quadratic
circuits that
compute the product of two $n \cross n$ matrices over finite fields.
In particular we obtain the following results:
\begin{enumerate}
\item{We show that the number of product gates in any {\it bilinear}
(or {\it quadratic}) circuit
that computes the product of two $n \cross n$ matrices over $\F(2)$ is at
least $3 n^2 - o(n^2)$.}
\item{We show that the number of product gates in any {\it bilinear} circuit
that computes the product of two $n \cross n$ matrices over $\F(p)$ is at
least $(2.5 + \frac{1.5}{p^3 -1})n^2 -o(n^2)$.}
\end{enumerate}
These results improve the former results of
\cite{Bshouty89,Blaser99} who proved lower bounds of  $2.5 n^2 - o(n^2)$.
\end{abstract}

\section{Introduction}

The problem of computing the product of two matrices is one of the most
studied computational problems:
We are given two $n \times n$ matrices
$x=(x_{i,j})$, $y=(y_{i,j})$, and we wish
to compute their product, i.e. there are $n^2$ outputs where the 
$(i,j)$'th output is
$$(x \cdot y)_{i,j} = \sum_{k=1}^{n}x_{i,k} \cdot y_{k,j} \;.$$

In 69' Strassen surprised the world by showing an upper bound of
$O(n^{\log_2{7}})$ \cite{Strassen69}. This bound was later improved and
the best upper bound today is $O(n^{2.376})$ \cite{CoppersmithWinograd90}
(see also \cite{Gathen88} for a survey).
The best lower bound is a  lower bounds of $2.5 n^2 - o(n^2)$,
on the number of products needed to compute the function 
\cite{Bshouty89,Blaser99}.
Thus the following problem is still open:
Can matrix product be computed by a circuit of size $O(n^2)$ ?

The standard computational model for computing polynomials is the model of
arithmetic circuits, i.e. circuits over the base $\{+, \; \cdot \}$ over some
field $F$. This is indeed the most general model, but for
matrix product two other models are usually considered,
{\it quadratic circuits} and {\it bilinear circuits}.
In the quadratic model we require that 
product gates are applied only on two linear functions. In the bilinear model 
we also require that product gates are applied only on two linear functions, 
but in addition we require that the first linear
function is linear in the variables of $x$ and that the second linear function
is linear in the variables of $y$.
These models are more restricted than the general model of
arithmetic circuits. However it is interesting to note that 
over infinite fields we can always assume w.l.o.g.
that any circuit for matrix product is a quadratic circuit
\cite{Strassen73a}. In addition we note that the best circuits that we
have today for matrix product are bilinear circuits.

In this paper we prove that any {\it quadratic} circuit that computes matrix
product over the field $\F(2)$ has at least $3n^2 - o(n^2)$ product gates, and
that any {\it bilinear} circuit for matrix product over the field $\F(p)$
must have at least $(2.5 + \frac{1.5}{p^3 -1})n^2 -o(n^2)$ product gates.

From now on we will use the notation $MP_n$ to denote the problem of
computing the product of two $n  \times n$ matrices.

\subsection{Known Lower Bounds}

In contrast to the major advances in proving upper bound, the attempts to
prove lower bounds on the size of bilinear circuits that compute $MP_n$
were less successful. Denote by $\sq$ and $\sbl$
the number of product gates in a smallest quadratic circuit for $MP_n$, and
in a smallest bilinear circuit for $MP_n$ respectively. We also denote by
$\tbl$ the total number of gates in a smallest bilinear circuit for $MP_n$.
In 78' Brocket and Dobkin proved that $\sbl \geq 2n^2 - 1$ over any field
\cite{BrocketDobkin78}. 
This lower bound was later generalized by Lafon and Winograd
to a lower bound on $\sq$ over any field \cite{LafonWinograd78}.
In 89' Bshouty showed that over $GF(2)$, $\sq \geq 2.5 n^2 - O(n \log n)$ 
\cite{Bshouty89}. Recently Bl\"{a}ser  proved a lower bound of $2n^2 + n -3$ on
$\sq$ over any field \cite{Blaser00}. 
In \cite{Blaser99} Bl\"{a}ser proved that 
$\sbl \geq 2.5 n^2 - 3n$ over any field.

In \cite{RazShpilka01} it is shown that any bounded depth circuit for $MP_n$,
over any field, has a super linear (in $n^2$) size.
Notice however, that the best known circuits
for $MP_n$ have depth $\Omega(\log n)$.

\subsection{Bilinear Rank}

An important notion that is highly related to the problem of computing
matrix product in bilinear circuits is the notion of bilinear rank.

A bilinear form in two sets of variables $x,y$ is a polynomial in the
variables of $x$ and the variables of  $y$,
which is linear in the variables of $x$ and linear 
in the variables of $y$. Clearly
each output of $MP_n$ is a bilinear form in $x=\{x_{i,j}\}, \; y=\{y_{i,j}\}$.
The bilinear rank of a set of bilinear forms
$\{\; b_1(x,y),\; \dots,\; b_m(x,y) \; \}$ is the
smallest number of rank 1 bilinear forms that span $b_1,\; \dots , \; b_m$,
where a rank 1 bilinear form is a product of a linear form in the
$x$ variables and a linear form in the $y$ variables.
We denote by $R_F(b_1,\; \cdots ,\; b_m)$ the bilinear rank of
$\{ \; b_1,\; \dots,\; b_m \; \}$ over the field $F$.
For further background see \cite{BuergisserClausenS97,Gathen88}.

We denote by $R_F(MP_n)$ the bilinear rank over $F$ of the $n^2$ outputs of
matrix product, i.e. it is the bilinear rank of the set
$\{ \sum_{k=1}^{n} x_{i,k} \cdot y_{k,j} \}_{i,j}$ over $F$.

The following inequalities are obvious (over any field).
\begin{itemize}
\item{$\sq \leq \sbl \leq 2 \sq$.}
\item{$R_F(MP_n) = \sbl$.}
\item{The following inequality is less obvious, but also not so hard to see.
$$\sbl \leq \tbl \leq \mathrm{poly}(\log n) \cdot \sbl \;.$$
I.e. up to polylogarithmic
factors, the number of product gates in a smallest bilinear
circuit for $MP_n$, over any field $F$, is equal to the total number of gates
in the circuit.}
\end{itemize}


\subsection{Results and Methods}

We prove that any {\it quadratic} circuit that computes $MP_n$
over the field $\F(2)$ has at least $3n^2 - o(n^2)$ product gates
(i.e. $\sq \geq 3n^2 - o(n^2)$ over $\F(2)$).
We also prove that over the field $\F(p)$ every {\it bilinear} circuit for
$MP_n$ must have at least
$(2.5 + \frac{1.5}{p^3 -1})n^2 -o(n^2)$ product gates
(i.e. $\sbl \geq (2.5 + \frac{1.5}{p^3 -1})n^2 -o(n^2)$ over $\F(p)$).
Both of these results actually hold for the bilinear rank as well.

The proof of the lower bound over $\F(2)$ is based on techniques from the
theory of linear codes. However, we cannot use known results from coding theory
in a straightforward way,
since we are not dealing with codes in which every two words are
distant, but rather with codes on matrices in which the distance between two
code words, of two matrices, is proportional to the rank of the difference of
the matrices. The reduction from circuits to
codes and the proof of the bound are given in section~\ref{gf2}.

The proof of the second bound is based on a lemma proved by Bl\"{a}ser in
\cite{Blaser99}. We prove that in the case of 
finite fields we can use the lemma 
with better parameters than those used by Bl\"{a}ser. 
This result is proved in section~\ref{gfp}.

\subsection{Organization of the paper}

In section~\ref{model} we present the models of bilinear circuits and 
quadratic circuits. In section~\ref{methods} we present some algebraic and
combinatorial tools that we need for the proofs of our lower bounds. 

In section~\ref{gf2} we introduce the notion of linear codes of matrices, 
and prove our lower bound on bilinear and quadratic
circuits that compute $MP_n$ over $\F(2)$. 
In section~\ref{gfp} we prove our lower bound on bilinear circuits
that compute $MP_n$ over $\F(p)$.


\section{Arithmetic Models}
\label{model}
In this section we present the models of quadratic circuits and bilinear
circuits. These
are the models for which we prove our lower bounds. We first give the
definition of a general arithmetic circuit.
An arithmetic circuit over a field $F$ is a directed acyclic graph as follows.
Nodes of in-degree 0 are
called inputs and are labeled with input variables.
Nodes of out-degree 0
are called outputs. Each edge is labeled with a constant from the field and
each node other than an input is labeled with one of the following
operations $\{ \; + \;,\; \cdot \;\}$, in the first case the
node is a plus gate and in the second case a
product gate.
The computation is done in the following way.
An input just computes the value of the variable
that labels it. Then, if
$v_1,\; \dots,\; v_k$ are the vertices that fan into $v$ then
we multiply the result of each $v_i$ with the value of the edge that connects
it to $v$. If $v$ is a plus gate we sum all the results, otherwise
$v$ is a product gate and
we multiply all the results.
Obviously the value computed by each node in the circuit is a polynomial 
over $F$ in the input variables.

We are interested in the problem of computing the product of two
$n \times n$ matrices, $MP_n$.
The input consists of two $n \times n$ matrices $x, \; y$.
The output is the matrix $x \cdot y$, i.e., there are $n^2$ outputs, and
the $(i,j)$'th output is:
$$ (x \cdot y)_{i,j} = \sum_{k=1}^{n} x_{i,k} \cdot y_{k,j} \;.$$
Each output $(x \cdot y)_{i,j}$ is hence a  bilinear form in $x$ and $y$.

Since each output of $MP_n$ is a bilinear form, it is natural to
consider bilinear arithmetic circuits for it. A bilinear arithmetic
circuit is an arithmetic circuit with the additional restriction
that product gates are applied only on two linear functions, 
one function is linear in the variables of $x$ and the other function
is linear in the variables of $y$. 
Thus, bilinear circuits have the following structure. First,
there are many plus gates computing linear forms in $x$ and
linear forms in $y$.
Then there is one level of product gates that compute bilinear forms, and
finally there are many plus gates that eventually compute the
outputs.
We will be interested in bounding from below the number of products in any
bilinear circuit for $MP_n$.
This model is more restricted than the general model of arithmetic circuits
but we note that all the known upper bounds
(over any field) for $MP_n$ are by bilinear circuits.

Another model that we will consider is the model of quadratic circuits.
A quadratic circuit is an arithmetic circuit with the additional restriction
that product gates are applied only on two
linear functions. Notice that the only difference between quadratic circuits
and bilinear circuits is that in the quadratic model the product gates
compute {\it quadratic forms} in $x,\; y$, whereas in the bilinear model the
product gates compute {\it bilinear forms} in $x,\; y$.
This model is more general than the model of bilinear circuits, but it
is still more restricted than the general model. 
However it is interesting to note that over infinite fields we can assume 
w.l.o.g.
that any arithmetic circuit for $MP_n$ is a quadratic circuit  
\cite{Strassen73a}.


\section{Algebraic and Combinatorial tools}
\label{methods}

In this section we present some algebraic and combinatorial tools that we 
will use.

The following  lemma is an extremely weak variant of the
famous Schwartz-Zippel lemma which shows that every non zero polynomial (non
zero as a formal expression)
over a large enough field has a non zero assignment in the field 
(see \cite{Schwartz80,Zippel79}).

\begin{lemma}
\label{schwartz}
Let $P$ be a polynomial of degree $d$ in $x_1,\; \dots, \; x_n$ over some
field $F$, such that $d < |F|$, and such that 
at least one of the coefficients of $P$ is not zero. 
Then we can find an assignment, $\rho \in F^n$, to the $x_i$'s, such that 
$P(\rho_1,\dots,\rho_n) \neq 0$. 
\end{lemma}

We say that two polynomials $p,q$ in $n$ variables are equivalent over a 
field $F$, if $p(x_1,\dots,x_n)=q(x_1,\dots,x_n)$
for any $x_1,\; \dots,\; x_n \in F$. We denote $p \equiv q$ if $p$ and $q$
are equivalent over $F$ (we omit $F$ from the notation as the field that we
deal with will be clear from the context).

\begin{lemma}
\label{solution}
Let $P$ be a polynomial of degree $d$ in the variables
$x_1,\; \dots,\; x_n$ over a field $F$.
If $P \not \equiv 0$ then we can find an assignment, $\rho \in F^n$, 
to the $x_i$'s  such that at most $d$ of the $\rho_i$'s 
get a nonzero value, and such that $P(\rho_1,\ldots,\rho_n) \neq 0$.
\end{lemma}

\begin{proof}
$P$ is equal (as a function) to a polynomial $\bar{P}$ 
in which the degree of each variable
is at most $|F|-1$. We call $\bar{P}$ the reduction of $P$.
Consider some monomial $M$ in $\bar{P}$ whose coefficient is not zero.
We assign all the variables that do not appear in $M$ to zero. 
The resulting polynomial (after the assignment), is a polynomial in the
variables of $M$, which is not the zero polynomial as it is a reduced 
polynomial which has a monomial with a non zero coefficient ($M$ of course). 
Therefore according to lemma~\ref{schwartz}
there is some assignment to the variables of $M$, 
that gives this polynomial a nonzero value. Therefor we have found an 
assignment which gives nonzero values only to the variables of $M$ 
(and there are at most $d$ such variables) under which $P \neq 0$.
\end{proof}

The following useful lemma, which is a straightforward implication of the
previous lemma, is the key lemma in most of our proofs.
The lemma deals with linear forms in $n^2$ variables. 
From now on we shall think about such
linear forms as linear forms in the entries of $n \times n$ matrices.

\begin{lemma}
\label{general_vanish}
Let $\mu_1,\; \dots,\; \mu_{n^2}$ be $n^2$ linearly independent linear forms
in $n^2$ variables over some field $F$.
Let $P$ be a polynomial of degree $d$ in $kn^2$ variables over $F$, 
i.e. we can view $P$ as a polynomial $P(x_1,...,x_k)$ in the entries of 
$k$ matrices, $x_1,...,x_k$, of size $n \times n$ each. 
Assume that $P \not \equiv 0$.
Then we can find $k$ matrices $a_1,\; \dots,\; a_k \in M_n(F)$ such that
$P(a_1,...,a_k) \neq 0$
and such that there exist $n^2 - d$ linear forms among
$\mu_1,...,\mu_{n^2}$'s that vanish on all the $a_i$'s. 
\end{lemma}

\begin{proof}
The idea of the proof is the following. Let $b_1,\; \dots,\; b_{n^2}$ be
the dual basis of $\mu_1,...,\mu_{n^2}$, i.e. it is a
basis of $M_n(F)$ satisfying $\forall i,j \;\; \mu_i(b_j) = \delta_{i,j}$.
We wish to find $k$ matrices, $a_1,...,a_k$, such that 
$P(a_1,...,a_k) \neq 0$, 
and such that there exist $b_{i_1},...,b_{i_d}$ that span all of them.
If we manage to find such matrices, then since the
$b_i$'s are the dual basis to the $\mu_i$'s we will get that $n^2 - d$ 
of the $\mu_i$'s vanish on $a_1,...,a_k$. The way to find
such matrices that are contained in the span of a small subset of the $b_i$'s,
is based on lemma~\ref{solution}.

So let $b_1,\; \dots,\; b_{n^2}$ be the dual basis to
$\mu_1,\; \dots,\; \mu_{n^2}$, i.e. 
$\forall i,j \;\; \mu_i(b_j)= \delta_{i,j}$.
We now change the variables of $P$. 
Let $\alpha_{i,j} \;\; j = 1...k, \; i=1...n^2$, 
be a set of $k n^2$ variables.
Denote $x_j = \sum_{i=1}^{n^2}\alpha_{i,j}b_i$.
Thus $P(x_1,...,x_k)$ can be viewed as a polynomial of degree $d$ in
the $k n^2$ variables $\alpha_{i,j}$. 
Therefore $P \not \equiv 0$ as a polynomial 
in the $\alpha_{i,j}$'s. 
Hence, according to lemma~\ref{solution} there exists an 
assignment, $\rho$, to the $\alpha_{i,j}$'s such that at most $d$ 
of them get a 
nonzero value. Define $a_j = \sum_{i=1}^{n^2}\rho_{i,j}b_i$.
Clearly $P(a_1,\ldots,a_k) \neq 0$.
Since at most $d$ of the $\rho_{i,j}$'s
got non zero values, we see that there are at most $d$ $b_i$'s such that 
all the $a_j$'s are linear combinations of them. Since the $b_i$'s are the dual
basis to $\mu_1,\; \dots,\; \mu_{n^2}$ we get that there are at least
$n^2 - d$ of the $\mu_i$'s that vanish on all the $a_j$'s. 
Therefore $a_1,\; \dots,\; a_k$
satisfy the requirements of the lemma.
\end{proof}

The next lemma will enable us to translate properties of matrices over
large fields of characteristic $p$ to properties of 
matrices (of higher dimension) over  $GF(p)$.

\begin{lemma}
\label{embed}
There exist an embedding, $\phi : GF(p^n) \hookrightarrow \Mnp$.
That is there exist a mapping $\phi : GF(p^n) \mapsto \Mnp$
such that
\begin{itemize}
\item{$\phi$ is a one to one linear transformation.}
\item{$\phi(1)= I$, where $I$ is the $n \times n$ identity matrix.}
\item{$\phi$ is multiplicative, i.e. $\forall x,y \in GF(p^n)$ we have that 
$\phi(xy)= \phi(x) \cdot \phi(y)$.}
\end{itemize}
This embedding also induces an embedding
$M_k(GF(p^n)) \hookrightarrow M_{nk}(GF(p))$.
\end{lemma}

This lemma is a standard tool in algebra, but for completeness we give the
proof.\\

\begin{proof}
$GF(p^n)$ is an $n$ dimensional vector space over $GF(p)$. Each
element $x \in GF(p^n)$ can be viewed as a linear transformation
$x:GF(p^n) \mapsto GF(p^n)$ in the following way: 
$$\forall y \in GF(p^n)  \;\; x(y) = x \cdot y \;.$$
Clearly this is a linear transformation of $GF(p^n)$ into itself, as a vector
space over $GF(p)$. Therefore, by picking a basis to $GF(p^n)$ we can represent
the linear transformation corresponding to each $x \in GF(p^n)$ by a matrix
$a_x \in M_n(GF(p))$. Thus, we have defined a mapping
$\phi :GF(p^n) \mapsto \Mnp$ such that $\phi(x)= a_x$,
and it is easy to verify that this mapping is an
embedding of $GF(p^n)$ into $M_n(GF(p))$. The way to generalize it to an
embedding of  $M_k(GF(p^n))$ into $M_{nk}(GF(p))$ is the following.
Let $a = (a_{i,j}) \in M_k(GF(p^n))$ be some matrix.
Every entree of $a_{i,j}$ of $a$,  is some element of $GF(p^n)$.
We can now replace $a_{i,j}$ with the matrix $\phi(a_{i,j})$.
Thus the resulting
matrix will be a $kn \times kn$ matrix whose entries are in $GF(p)$. Again it
is easy to verify that this is indeed an embedding of
$M_k(GF(p^n))$ into $M_{nk}(GF(p))$.
\end{proof}

In addition to the algebraic lemmas we also need the following  combinatorial
tools.

\begin{dfn}
Let $F$ be a field, and let $v, \; u$ be two vectors in $F^m$.
We denote by $\mathrm{\bf weight}(v)$ 
the number of nonzero coordinates of $v$.
Let $\mathrm{\bf d_H}(v,u) = \mathrm{\bf weight}(v-u)$, 
i.e. $\mathrm{\bf d_H}(v,u)$
is the number of coordinates on
which $u$ and $v$ differ. 
$\mathrm{\bf d_H}(v,u)$ is also known as the Hamming distance of
$u$ and $v$. We also denote by 
$\mathrm{\bf agree}(u,v)$ the number of coordinates
on which $u$ and $v$ are equal, i.e. 
$\mathrm{\bf agree}(u,v) = m-\mathrm{\bf d_H}(v,u)$.
\end{dfn}

The next lemma shows that if a vector space contains a set of vectors
such that every pair/triplet of them don't agree on many coordinates 
(i.e. their Hamming distance is large) then it is of large dimension.
There are numerous similar lemmas in coding theory, and in particular
the first part of our lemma is the famous Plotkin bound 
(see \cite{vanLint92}).

\begin{lemma}
\label{intersect}
\begin{enumerate}
\item{In every set of $k$ vectors in $GF(p)^{t}$, 
such that $p<k$, there are two
vectors that agree on at least $(\frac{t}{p}  - \frac{t}{k})$ coordinates.}
\item{In every set of $k$ vectors in $GF(p)^{t}$, 
such that $2p<k$, there are three
vectors that agree on at least 
$(\frac{t}{p^2}  - \frac{3t}{pk})$ coordinates.}
\end{enumerate}
\end{lemma}

\begin{proof}
We begin by proving the first claim.
Let $v_1,\ldots,v_k$ be $k$ vectors in $GF(p)^{t}$. We are going
to estimate $\sum_{i < j}\mathrm{\bf agree}(v_i,v_j)$ in two different ways.
On the one hand
this sum is at most ${k \choose 2}$ times the maximum of
$\mathrm{\bf agree}(v_i,v_j)$.
On the other hand consider a certain coordinate.
For  every $\alpha \in GF(P)$ denote by $n_{\alpha}$ 
the number of vectors among the $v_i$'s that are
equal to $\alpha$ on this coordinate.
Clearly $\sum_{\alpha = 0}^{p-1}n_{\alpha} = k$. The contribution
of this coordinate to $\sum_{i < j}\mathrm{\bf agree}(v_i,v_j)$ is exactly
$\sum_{\alpha = 0}^{p-1} {n_{\alpha} \choose 2}$. By convexity
$$\sum_{\alpha = 0}^{p-1} {n_{\alpha} \choose 2} \geq p \cdot
\frac{1}{2} \cdot \frac{k}{p} \left( \frac{k}{p} -1 \right) = \frac{k(k-p)}{2p} \;.$$
We get that
$${k \choose 2} \cdot \max_{i<j}(\mathrm{\bf agree}(v_i,v_j)) \geq$$
$$\sum_{i < j}\mathrm{\bf agree}(v_i,v_j) \geq t \cdot \frac{k(k-p)}{2p} \;.$$
Therefore
$$\max_{i<j}(\mathrm{\bf agree}(v_i,v_j))
\geq \frac{t}{p} \cdot \frac{k-p}{k-1} \geq
\frac{t}{p} \cdot \frac{k-p}{k} = \frac{t}{p} - \frac{t}{k} \;.$$
The proof of the second claim is similar. We give two different
estimates to $\sum_{i < j < l}\mathrm{\bf agree}(v_i,v_j,v_l)$
(the number of coordinates on which $v_i,\; v_j$, and $v_l$ are the same).
In the same manner as before we get that
$$\max_{i< j < l}(\mathrm{\bf agree}(v_i,v_j,v_l))
\geq \frac{t}{p^2}  \cdot \frac{k-p}{k-1} \cdot \frac{k-2p}{k-2} \geq
\frac{t}{p^2} - \frac{3t}{pk} \;. $$
\end{proof}

\begin{cor}
\label{dist}
If $\zo{t}$ contains $k$ vectors $v_1,\; \dots, \;v_k$,
such that $2 < k$ and
$\forall i \neq j \; \mathrm{\bf d_H}(v_i,v_j) \geq N$, then
$t \geq 2N - 4\frac{N}{k+2}$.
\end{cor}

\begin{proof}
According to lemma~\ref{intersect} there are two vectors, w.l.o.g. $v_1$
and $v_2$, 
such that $\mathrm{\bf agree}(v_1,v_2) \geq \frac{t}{2}-\frac{t}{k}$.
Since $\mathrm{\bf d_H}(v_1,v_2)=t - \mathrm{\bf agree}(v_1,v_2)$ we get that
$$t - (\frac{t}{2}-\frac{t}{k}) \geq \mathrm{\bf d_H}(v_1,v_2) \geq N $$
and the result follows.
\end{proof}



\section{Lower bound over GF(2)}
\label{gf2}

In this section we prove our main theorems.

\begin{theorem}
\label{main_rank}
$\sbl \geq 3 n^2 - O(n^{\frac{5}{3}}) $ (in other words $R_{GF(2)}(MP_n) \geq
3 n^2 - O(n^{\frac{5}{3}}) $).
\end{theorem}

The second theorem that we shall prove is a lower bound for quadratic circuits.

\begin{theorem}
\label{main_qu}
$\sq \geq 3 n^2 - O(n^{\frac{5}{3}}) $. I.e.  the number of product gates in 
any quadratic circuit that computes the product
of two $n \times n$ matrices over $GF(2)$ is at least
$3 n^2 - O(n^{\frac{5}{3}}) $.
\end{theorem}

Clearly theorem~\ref{main_qu} imply theorem~\ref{main_rank}, but
we first prove of theorem~\ref{main_rank} 
as it is more intuitive and simple. 
We begin by introducing the notion of linear codes of matrices.


\subsection{Linear Codes of Matrices}

\begin{dfn}
A linear code of matrices is a mapping,
$$\Gamma: \Mn \mapsto \zo{m} \;,$$
(for some $m$) with the following properties:
\begin{itemize}
\item{ $\Gamma$ is linear.}
\item{For any matrix $a$, 
$\mathrm{\bf weight}(\Gamma(a)) \geq n \cdot \mathrm{rank}(a)$.}
\end{itemize}
\end{dfn}

From the linearity of $\Gamma$ and the requirement
on $\mathrm{\bf weight}(\Gamma(a))$ we get the following corollary.

\begin{cor}
$\Gamma$ is a one to one mapping, and
for any two matrices $a$ and $b$, 
$\mathrm{\bf d_H}(\Gamma(a),\Gamma(b)) \geq n \cdot \mathrm{rank}(a-b)$.
\end{cor}

The following theorem shows that the dimension of the range of
any linear code of matrices is large (i.e. $m$ must be large).

\begin{theorem}
\label{codes}
Let $\Gamma: \Mn \mapsto \zo{m}$ be a linear code of matrices,
then $m \geq 3 n^2 - O(n^{\frac{5}{3}}) \;.$
\end{theorem}

\begin{proof}
Denote 
$$\Gamma(a) = (\; \mu_1(a),\; \dots,\; \mu_m(a) \;) \;.$$
The proof is based on the following lemma that shows
that  we can find $k=n^{\frac{1}{3}}$ matrices,
$a_1,\; \dots,\; a_k \in \Mn$, with the following properties.
\begin{itemize}
\item{$\forall i \neq j \;,\; a_i - a_j$ is an invertible matrix.}
\item{There are $n^2 - {k \choose 2}n$ linear forms among the $\mu_i$'s
that vanish on all the $a_i$'s.}
\end{itemize}

We state the lemma for every $k < 2^n$ but we apply it only
to $k = n^{\frac{1}{3}}$.

\begin{lemma}
\label{k-vanish}
For every $n,k$ such that $k < 2^n$, and any
$\mu_1, \; \dots, \; \mu_{n^2}$ linearly independent linear forms in $n^2$ 
variables, over $GF(2)$,
there are $k$ matrices, $a_1,  \dots,  a_k \in \Mn$,
such that for every $i \neq j$,
$a_i - a_j$ is an invertible matrix, and such that 
$n^2 - {k \choose 2}n$ of the $\mu_i$'s vanish on them.
\end{lemma}

\begin{proof}
Consider the following polynomial $P$ in $k$ matrices:
$$P(a_1,\ldots,a_k) = \mathrm{determinant}\left 
(\prod_{i < j}(a_i - a_j) \right) \;.$$
Clearly a set of $k$ matrices $a_1, \ldots, a_k$ satisfy
$P(a_1,\ldots,a_k) \neq 0$
iff all the matrices $a_i - a_j$ are invertible. 
In addition, it is easy to see that
$deg(P) = {k \choose 2}n $.
Therefore if we show that $P \not \equiv 0$ over $GF(2)$,
then according to lemma~\ref{general_vanish} we will
get what we wanted to prove.

In order to show that $P \not \equiv 0$ we just have to prove the existence 
of $k$ matrices, such that the difference of every two of them is invertible.
Lemma~\ref{embed} assures us that we can embed the field $GF(2^n)$ into $\Mn$.
Denote this embedding by $\Phi:GF(2^n) \hookrightarrow \Mn$.
We take $k$ distinct elements in $GF(2^n)$, $x_1,\; \dots, \; x_k$.
Their images, $\Phi(x_1), \; \dots,\; \Phi(x_k)$, are matrices in $\Mn$
such that the difference of every two of them,
$\Phi(x_i)- \Phi(x_j) = \Phi(x_i - x_j)$, is an invertible matrix.
This is because the $x_i$'s are distinct (i.e. $x_i - x_j \neq 0$), 
and every nonzero element in $\F(2^n)$ is invertible.
Thus, $\Phi(x_1), \; \dots,\; \Phi(x_k)$ are exactly
the $k$ matrices that we were looking for. This concludes the proof of the
lemma.
\end{proof}

We proceed with the proof of the theorem.
Let $k = n^{\frac{1}{3}}$.
Since $\Gamma$ is a one to one mapping,
there are $n^2$ independent
linear forms among $\mu_1,\; \ldots, \;\mu_m$.
Therefore we can use lemma~\ref{k-vanish} and get that
there are $k$ matrices $a_1,\; \ldots,\; a_k$
such that for every $i \neq j$
$a_i - a_j$ is invertible, and such that, w.l.o.g.,
$\mu_{m-r+1},\; \dots,\; \mu_m$ vanish on $a_1,\; \dots,\; a_k$ for
some
$r \geq n^2 - {k \choose 2}n \geq n^2 - n^{\frac{5}{3}}$.

Since the last $r$ linear forms vanish on all the $a_i$'s, we are going to
restrict our attention only to the first $m-r$ linear forms. So from now on we
only consider $\Gamma(a_i)$ restricted to its first $m-r$ coordinates.

Since each of the
differences, $a_i - a_j$ ($\forall i \neq j$), 
is an invertible matrix, we get that \linebreak
$\mathrm{\bf d_H}(\Gamma(a_i),\Gamma(a_j)) \geq n^2$.
Thus, $\Gamma(a_1),\; \dots,\; \Gamma(a_k)$ are $k$ vectors contained in
$\zo{m-r}$ (we consider only their first $m-r$ coordinates !) such that the
hamming distance of every pair of them is at least $n^2$. 
Therefore according to corollary~\ref{dist} we get that
$$m-r \geq 2n^2 - 4\frac{n^2}{k+2} \;.$$
Since $r \geq n^2 - n^{\frac{5}{3}}$ and $k = n^{\frac{1}{3}}$, we get that
$$ m \geq 3 n^2 - O(n^{\frac{5}{3}})$$
which is what we wanted to prove.
This concludes the proof of the theorem.
\end{proof}


\subsection{Proof of Theorem~\ref{main_rank}}

Assume that $\sbl = m$.
Let $C$ be a smallest bilinear circuit for $MP_n$.
Let 
$$\mu_1(x) \cdot \eta_1(y),\; \dots, \; \mu_m(x) \cdot \eta_m(y)$$
be the $m$ bilinear forms computed in the product gates of $C$.
We will show that these bilinear forms define in a very
natural way a  code on $\Mn$.
The code thus defined, will have the property that
the dimension of the space into which the code maps $\Mn$
is exactly $m$.
Thus, according to theorem~\ref{codes} we will get that
$m \geq 3n^2 - O(n^{\frac{5}{3}})$, which is what we wanted to prove.

So we begin by defining a mapping from $\Mn$ to $\zo{m}$.
Let $\Gamma: \Mn \mapsto \zo{m}$ be the following mapping.
$$\Gamma(x)= (\mu_1(x),\; \dots,\; \mu_m(x)) \;.$$
Notice that we ignore the $\eta_i$'s in the definition of $\Gamma$.
The next lemma shows that $\Gamma$ is a linear code of matrices.

\begin{lemma}
$\Gamma$ is a linear transformation with the property that for 
every matrix $x \in \Mn$, $\mathrm{\bf weight}(\Gamma(x)) \geq
n \cdot \mathrm{rank}(x)$.
\end{lemma}

\begin{proof}
Clearly $\Gamma$ is a linear transformation from
$\Mn$ to $\zo{m}$. So we only have to prove the claim
about the weights.
Let $x$ be a matrix of rank $r$. Assume w.l.o.g. that
$\mu_1(x)= \dots = \mu_k(x)=1$ and that $\mu_{k+1}(x)= \dots = \mu_m(x)=0$,
i.e. $\mathrm{\bf weight}(\Gamma(x)) = k$.
We shall show that $k \geq nr$.
For every $y \in \Mn$, the $n^2$ entries of $x \cdot y$ are 
functions of 
$\mu_1(x) \cdot \eta_1(y),\; \dots, \; \mu_m(x) \cdot \eta_m(y)$.
Since $\mu_{k+1}(x)= \dots = \mu_m(x)=0$, 
we get that $x \cdot y$ is a function of
$\eta_1(y),\; \dots, \; \eta_k(y)$.
Therefore there are at most $2^k$ different matrices of the form $x\cdot y$.
Since $\mathrm{rank}(x)=r$ we get that there are exactly
$2^{nr}$ different matrices of the form $x \cdot y$.
Therefore $k \geq nr$. This concludes the proof of the lemma.
\end{proof}

Therefore $\Gamma$ is a linear code of matrices, so according to 
theorem~\ref{codes} we get that
$m \geq 3n^2 - O(n^{\frac{5}{3}}) $
which is what we wanted to prove.
This concludes the proof of theorem~\ref{main_rank}.
{\hfill $\clubsuit$}


\subsection{Proof of Theorem~\ref{main_qu}}
\label{gf2_app}



As in the proof of theorem~\ref{main_rank} we 
will show that every quadratic circuit for $MP_n$,
defines a code on $\Mn$.
The code thus defined, will have the property that $m$
(i.e the dimension of the space into which the code maps $\Mn$)
is exactly the number of product gates in the circuit.
Thus, according to theorem~\ref{codes} we will get that
$m \geq 3n^2 - O(n^{\frac{5}{3}})$, which is what we wanted to prove.

Let $C$ be a quadratic circuit for $MP_n$.
Assume that the product gates of $C$ compute the quadratic forms
$\mu_1(x,y) \cdot \eta_1(x,y),\; \dots, \; \mu_m(x,y) \cdot \eta_m(x,y)$.
Thus, each of the outputs $(x \cdot y)_{i,j}$ can be written as a sum of
these quadratic forms:
$$(x \cdot y)_{i,j} = \sum_{k=1}^{m} \alpha_{i,j}^{(k)} \cdot
\mu_k(x,y) \cdot \eta_k(x,y) \;,$$
where $\alpha_{i,j}^{(k)} \in \zozo$.

We would like to have a proof similar to the proof of theorem~\ref{main_rank}. 
In that proof we defined a code of matrices using
the linear transformation $\mu_1,...,\mu_m$. Unfortunately this method
will fail here as $\mu_i$ is a linear function in both the variables of $x$
and the variables of $y$ and not just in the variables of $x$ as in
the proof of theorem~\ref{main_rank}. In order to overcome this obstacle
we introduce a new set of variables $z=\{z_{i,j}\}_{i,j=1...n}$. 
We think about $z$ as an $n \times n$ matrix. 
Define the following $m$ linear forms in $z$:
$$\gamma_k(z) = \sum_{i,j} \alpha_{i,j}^{(k)} z_{i,j} \;,\;\; k=1,\dots,m \;.$$
We get that
$$ \sum_{k=1}^{m} \mu_k(x,y) \cdot \eta_k(x,y) \cdot \gamma_k (z) =$$
$$\sum_{k=1}^{m} (\sum_{i,j} z_{i,j} \cdot \alpha_{i,j}^{(k)}) \cdot
\mu_k(x,y) \cdot \eta_k(x,y) = $$
\begin{equation}
\label{trace_eq}
\sum_{i,j} z_{i,j} \sum_{k=1}^{m}  \alpha_{i,j}^{(k)} \cdot
\mu_k(x,y) \cdot \eta_k(x,y) = 
\end{equation}
$$\sum_{i,j} z_{i,j} \cdot (x \cdot y)_{i,j} =
\mathrm{\bf trace} (x \cdot y \cdot z^{t}) \;,$$
where $(z^t)_{i,j} = z_{j,i}$.
The computation that we just performed shows that the $\gamma_k$'s that
we introduced are quite natural. We also notice that $z$ plays the same 
role in $\mathrm{\bf trace} (x \cdot y \cdot z^{t})$
as $x$ and $y$. These observations motivate us to try to repeat the
proof of theorem~\ref{main_rank} using the $\gamma_k$'s instead of the
$\mu_i$'s.

So define a linear mapping $\Gamma: \Mn \mapsto \zo{m}$ by
$$\Gamma(z) = (\gamma_1(z),\; \dots,\; \gamma_m(z))\;.$$
The following lemma shows that $\Gamma$ is indeed a linear code of matrices.

\begin{lemma}
$\Gamma$ is a linear mapping and it has the property that
for every matrix $z$, $\mathrm{\bf weight}(\Gamma(z)) 
\geq n \cdot \mathrm{rank}(z)$.
\end{lemma}

\begin{proof}
Clearly $\Gamma$ is a linear mapping. 
So we only have to prove the claim about the weights.
Let $z_0$ be a matrix of rank r, and assume w.l.o.g. that
$\gamma_1(z_0)= \ldots = \gamma_k(z_0) = 1$ and
$\gamma_{k+1}(z_0)= \ldots = \gamma_m(z_0) = 0$.
We wish to prove that $k \geq nr$.
From equation~\ref{trace_eq} we get that 
$$ \mathrm{\bf trace}(x \cdot y \cdot {z_0}^t) =
\sum_{i=1}^{k} \mu_i(x,y) \cdot \eta_i(x,y) \;.$$
We now consider the discrete derivatives of this equation. Let $e_{i,j}$ be
the matrix of all zeros but 1 in the $(i,j)$'th place.
Define
$$\frac{\partial}{\partial x_{i,j}}\mathrm{\bf trace}(x \cdot y \cdot {z_0}^t)
\stackrel{\mathrm{def}}{=}$$
$$\mathrm{\bf trace}((x+e_{i,j}) \cdot y \cdot {z_0}^t) -
\mathrm{\bf trace}(x \cdot y \cdot {z_0}^t) \;.$$
On the one hand
$$\mathrm{\bf trace}((x+e_{i,j}) \cdot y \cdot {z_0}^t) -
\mathrm{\bf trace}(x \cdot y \cdot {z_0}^t) =$$
$$\mathrm{\bf trace}(e_{i,j} \cdot y \cdot {z_0}^t) = 
(z_0 \cdot y^t)_{i,j} \;.$$
On the other hand we have that
$$ \mathrm{\bf trace}((x+e_{i,j}) \cdot y \cdot {z_0}^t) -
\mathrm{\bf trace}(x \cdot y \cdot {z_0}^t)=$$
$$\sum_{i=1}^{k} (\mu_i(x+e_{i,j},y) \cdot \eta_i(x+ e_{i,j},y) -
\mu_i(x,y) \cdot \eta_i(x,y)) =$$
\begin{equation}
\label{trace_dif}
\sum_{i=1}^{k} (\mu_i(e_{i,j},0) \cdot \eta_i(x,y) +
\mu_i(x,y) \cdot \eta_i(e_{i,j},0)) +
\end{equation}
$$+ \sum_{i=1}^{k} \mu_i(e_{i,j},0) \cdot
\eta_i(e_{i,j},0) \;,$$
where the last equality follows from the linearity of the $\mu_i$'s and the
$\eta_i$'s.
Since $(z_0 \cdot y^t)_{i,j}$ is a linear form in $y$, we actually get that
$$(z_0 \cdot y^t)_{i,j} = \frac{\partial}{\partial x_{i,j}}\mathrm{\bf trace}
(x \cdot y \cdot {z_0}^t)=$$ 
$$\sum_{i=1}^{k} (\mu_i(e_{i,j},0) \cdot \eta_i(x,y) +
\mu_i(x,y) \cdot \eta_i(e_{i,j},0))$$
$$\subset \mathrm{span}(\mu_i(x,y),\; \eta_i(x,y))\;$$
(since $(z_0 \cdot y^t)_{i,j}$ is a linear form the third summand of
equation~\ref{trace_dif} sums to $0$).  
In the same manner we define
$$\frac{\partial}{\partial y_{i,j}}\mathrm{\bf trace}(x \cdot y \cdot {z_0}^t)
\stackrel{\mathrm{def}}{=}$$
$$\mathrm{\bf trace}(x \cdot (y+ e_{i,j}) \cdot {z_0}^t) -
\mathrm{\bf trace}(x \cdot y \cdot {z_0}^t) \;.$$
We get that
$$(x^t \cdot z_0)_{i,j} = \frac{\partial}{\partial y_{i,j}}\mathrm{\bf trace}
(x \cdot y \cdot {z_0}^t) = $$
$$\sum_{i=1}^{k} (\mu_i(0,e_{i,j}) \cdot \eta_i(x,y) +
\mu_i(x,y) \cdot \eta_i(0,e_{i,j})) $$
$$\subset \mathrm{span}(\mu_i(x,y),\; \eta_i(x,y))\;.$$
Denote by $\mathrm{PD}$ the set of    
all the discrete partial derivatives
$$\left\{ \frac{\partial}{\partial x_{i,j}}\mathrm{\bf trace}
(x \cdot y \cdot {z_0}^t),
\frac{\partial}{\partial y_{i,j}}\mathrm{\bf trace}(x \cdot y \cdot {z_0}^t)
\right \} _{i,j} \;.$$
We just proved that $\mathrm{PD}$ is contained in the linear span of
$$\{ \; \mu_i(x,y),\; \eta_i(x,y) \; \}_{i=1}^{k} \;,$$
in the vector space of all linear forms in $x, \; y$.
Therefore
$$\mathrm{dim}( \mathrm{span} (\mathrm{PD})) \leq $$
\begin{equation}
\label{spanleq}
\leq \mathrm{dim}( \mathrm{span} \{ \; \mu_i(x,y),\;
\eta_i(x,y) \; \}_{i=1}^{k}) \leq 2k \;.
\end{equation}
We also showed that
$$\left\{ \frac{\partial}{\partial x_{i,j}}\mathrm{\bf trace}
(x \cdot y \cdot {z_0}^t),
\frac{\partial}{\partial y_{i,j}}\mathrm{\bf trace}(x \cdot y \cdot {z_0}^t)
\right \} _{i,j} = $$
$$= \left\{ \; (x^t \cdot z_0)_{i,j}, \; (z_0 \cdot y^t)_{i,j} \;
\right\}_{i,j} \;.$$
Therefor, using our assumption that $\mathrm{rank}(z_0)=r$, we get that
$$\mathrm{dim}( \mathrm{span}(\mathrm{PD}))=$$
\begin{equation}
\label{spangeq}
= \mathrm{dim}( \mathrm{span}\left\{ \; (x^t \cdot z_0)_{i,j}, \;
(z_0 \cdot y^t)_{i,j} \right \}_{i,j}) = 2nr \;.
\end{equation}
Combining equations~\ref{spanleq} and \ref{spangeq} we get that $2k \geq 2nr$.
\end{proof}

Theorem~\ref{main_qu} 
now follows from applying theorem~\ref{codes} on the linear code
of matrices $\Gamma$. 
{\hfill $\clubsuit$}


\section{Other Finite Fields}
\label{gfp}

In this section we prove the following theorem.

\begin{theorem}
\label{main_finite}
The number of product gates in any bilinear circuit that computes
the product of two $n \times n$ matrices over $GF(p)$ is at least
$(2.5 + \frac{1.5}{p^3 -1})n^2 - O(n^{\frac{7}{4}})$
(i.e. $\sbl \geq (2.5 + \frac{1.5}{p^3 -1})n^2 - O(n^{\frac{7}{4}})$
over $GF(p)$).
\end{theorem}

Let $C$ be a bilinear circuit for $MP_n$ over $\F(p)$.
Assume that $\mu_1(x) \cdot \eta_1(y),\; \dots, \; \mu_m(x) \cdot \eta_m(y)$
are the bilinear forms computed in the product gates of $C$.
The following lemma of Bl\"{a}ser is the main tool in the proof of the
theorem.

\begin{lemma}\cite{Blaser99}
\label{bla}
Let $[a,b]=ab-ba$.
If there are two matrices $a,\; b$ such that $[a,b]$ is an invertible
matrix, and such that there are $t$ linear forms among 
$\mu_1,\;  \ldots,\; \mu_m$ such that each of them vanish on  
$I,\; a,\; b$ 
then $$m \geq t + 1.5 n^2 \;.$$
\end{lemma}

We are going to prove that we can find $a,b$   such that
$(1-\frac{1}{p^3}) n^2 + \frac{m}{p^3} - O(n^{\frac{7}{4}})$ linear forms
among $\mu_1,\; \dots, \; \mu_m$ vanish on $I,\; a,\; b$ ,
and such that $[a,b]$ is invertible. \\

\begin{proofof}
{\bf Theorem~\ref{main_finite}:}
We begin by proving that (w.l.o.g.) many of the $\mu_i$'s vanish on $I$.
The following lemma shows that we can always find an invertible matrix 
such that many of the $\mu_i$'s vanish on it. As before we assume that
$\mu_1,\ldots,\mu_{n^2}$ are independent linear forms.

\begin{lemma}
\label{inv}
There exists an invertible matrix $c$, such that at least
$(1-\frac{1}{p})n^2 + \frac{m}{p} - O(n^{\frac{5}{3}})$  of
the $\mu_i$'s vanish on it, where $n^2 -O(n^{\frac{5}{3}})$ of the $\mu_i$'s
that vanish on it are among $\mu_1,\; \ldots,\; \mu_{n^2}$.
\end{lemma}

\begin{proof}
An analog of lemma~\ref{k-vanish} over $GF(p)$ guarantees that we can
find $k=n^{\frac{1}{3}}$ matrices, $a_1,\; \ldots,\; a_k \in \Mnp$,
such that for $i \neq j$
$a_i - a_j$ is invertible, and such that
$n^2 - {k \choose 2}n$ linear forms among $\mu_1,\; \ldots,\; \mu_{n^2}$
vanish on all of them.
Denote $r=n^2 - {k \choose 2}n$. And assume
w.l.o.g. that $\mu_1,\; \dots, \; \mu_r$ vanish on all the $a_i$'s.

Let us consider the following k vectors in $GF(p)^{m- r}$:
$$\Gamma(a_i) \stackrel{\mathrm{def}}{=} 
(\mu_{r +1}(a_i),\ldots,\mu_m(a_i)) \;,\; i=1 \ldots k \;.$$
As in the proof of theorem~\ref{main_rank}, we get that since
$\forall i \neq j \; a_i -a_j$ is an invertible matrix, then
$\mathrm{\bf d_H}(\Gamma(a_i),\Gamma(a_j)) \geq n^2$.
According to lemma~\ref{intersect}, two of these vectors agree on at least
$\frac{m-r}{p} - \frac{m-r}{k}$ coordinates.
Assume that $\Gamma(a_1)$ and $\Gamma(a_2)$ are these vectors.
Denote $c=a_1 - a_2$.
We have that $c$ is an invertible matrix, such that the
first $r= n^2 - {k \choose 2}n$ linear forms
(which are independent) vanish on it, and such that all the linear forms
that $\Gamma(a_1)$ and $\Gamma(a_2)$ agree on, vanish on it as well.
Therefore there are at least
$$r + \frac{m-r}{p} - \frac{m-r}{k}$$
linear forms that vanish on $c$.
Since $r= n^2 - {k \choose 2}n$, and $k = n^{\frac{1}{3}}$,
we get that at least
$$(1-\frac{1}{p})n^2 + \frac{m}{p} - O(n^{\frac{5}{3}})$$
linear forms vanish on $c$, 
$n^2 - O(n^{\frac{5}{3}})$ of them are among
$\mu_1,\; \dots,\; \mu_{n^2}$ (we assume for simplicity that
$n^2 < m < 10n^2$, as it will not change the results).
This completes the proof of the lemma.
\end{proof}

The lemma doesn't tell us who $c$ is, but using the {\it sandwiching method}
we can assume that $c=I$:
We know that $x \cdot y$ is computed using the bilinear forms
$$\mu_1(x) \cdot \eta_1(y),\; \dots,\; \mu_m(x) \cdot \eta_m(y) \;.$$
We now do the following trick:
$x \cdot y = (x \cdot c) \cdot (c^{-1} \cdot y)$, therefore $x \cdot y$
can be computed using the bilinear forms
$$\tilde{\mu_1}(x) \cdot \tilde{\eta_1}(y),\; \dots,\; \tilde{\mu_m}(x) \cdot
\tilde{\eta_m}(y) \;,$$
where
$$\tilde{\mu_i}(x) \stackrel{\mathrm{def}}{=} \mu_i(x \cdot c) \;\;
\mathrm{and} \;\;
\tilde{\eta_i}(y) \stackrel{\mathrm{def}}{=} \eta_i(c^{-1} \cdot y) \;.$$
Thus, if $\mu_i(c) = 0$ then we get that
$\tilde{\mu_i}(I) = \mu_i(I \cdot c) = 0$.
This trick is called sandwiching, for further background see \cite{Blaser99,Groote78}.

So by combining the sandwiching method and lemma~\ref{inv}
we get that we can assume w.l.o.g. that 
$(1-\frac{1}{p})n^2 + \frac{m}{p} - O(n^{\frac{5}{3}})$ of the $\mu_i$'s, 
where $n^2 -O(n^{\frac{5}{3}})$ of them are among 
$\mu_1,\; \dots,\; \mu_{n^2}$, vanish on $I$.
The next lemma now assures us that we can find two
matrices $a,b$ that satisfy the requirements of lemma~\ref{bla}.

\begin{lemma}
\label{2-commutator}
There are two matrices $a,\; b$ such that $[a,b]$ is an invertible matrix and
such that at least
$(1-\frac{1}{p^3})n^2 + \frac{m}{p^3} - O(n^{\frac{7}{4}})$ of the $\mu_i$'s
vanish on $I,\; a,\; b$.
\end{lemma}

\begin{proof}
The proof of this lemma is similar to the proof of lemma~\ref{inv}. Let
$k=n^{\frac{1}{4}}$.
The following lemma shows that we can find $k$ matrices such that many of 
the $\mu_i$'s vanish on all of them and such that among their differences 
there are matrices satisfying the requirements of lemma~\ref{bla}.

\begin{lemma}
\label{k-commutator}
For every $n,k$, such that $p^{\frac{n}{2}} > 4{k \choose 3}$, and any
$\mu_1,\; \dots,\; \mu_{n^2}$ linearly independent linear forms, in $n^2$
variables, over $GF(p)$,
there are $k$ matrices, $a_1,\; \dots,\; a_k$,
such that $\forall i < j<l$,
$[a_i - a_l, a_j - a_l]$ is invertible, and such that 
$n^2 - 2{k \choose 3}n$ of the $\mu_i$'s vanish on all the $a_i$'s.
\end{lemma}

\begin{proof}
Again we use lemma~\ref{general_vanish}. Let $P$ be the following polynomial.
$$P(a_1,...,a_k)= \mathrm{determinant} 
\left(\prod_{i < j < l}[a_i - a_l,a_j - a_l] \right) .$$
Clearly $deg(P) = 2 {k \choose 3} n$ (as a polynomial in the entries of
the $a_i$'s). Therefore if we will prove that $P \not \equiv 0$, i.e. that
there exist $k$ matrices on which $P$ is not zero, then according to
lemma~\ref{general_vanish} we are done. This is guaranteed by the following
lemma.

\begin{lemma}
If $p^{\frac{n}{2}} > 4{k \choose 3}$ then
there exist $k$ matrices in $\Mnp$, $a_1,\; \ldots,\; a_k$ such that
$\forall i <j<l$, $[a_i - a_l, a_j - a_l]$ is invertible.
\end{lemma}

We prove the lemma only for $n$ even. Clearly this will not affect 
theorem~\ref{main_finite}, as the lower bound for odd $n$ follows from the 
lower bound for even $n$. \\

\begin{proof}
Consider the following polynomial in ${\bf 4k}$ 
variables (i.e. it is a polynomial
in $k$ matrices over $\bf{M_2(GF(p))}$! ).
$$Q(x_1,...,x_k)= \mathrm{determinant}\left(\prod_{i<j<l}[a_i -a_l,a_j-a_l]
\right) \;.$$
$Q$ is a polynomial of degree $d= 4 {k \choose 3}$ over $GF(p)$, in the
entries of the $a_i$'s.
Clearly $Q$ is not the zero polynomial (as it is a product of non zero
polynomials). Consider the field $F=GF(p^{\frac{n}{2}})$. 
Since $d< |F|$ we get by lemma~\ref{schwartz}
that there are $k$ matrices $\rho_1,...,\rho_k \in M_2(F)$
such that $Q(\rho_1,...,\rho_k) \neq 0$. 
That is, $\forall i<j<l \; [\rho_i-\rho_l,\rho_j-\rho_l]$ 
is an invertible matrix.
According to lemma~\ref{embed} we can embed $M_2(F)$ in $\Mnp$.
Therefore there are $k$ matrices in $\Mnp$ satisfying
$\forall i<j<l \; [a_i-a_l,a_j-a_l]$ is an invertible matrix,
which is what we wanted to prove.
\end{proof}

This concludes the proof of lemma~\ref{k-commutator}.
\end{proof}

We proceed with the proof of lemma~\ref{2-commutator}.
We now restrict our attention to the linear forms among
$\mu_{n^2 +1},\; \ldots,\; \mu_m$ that vanish on $I$.
We shall prove that three of the matrices guaranteed by 
lemma~\ref{k-commutator} agree on many of these linear forms
(more formally on $\frac{m- n^2}{p^3} - O(n^{\frac{7}{4}})$ of them).
Thus, if $a_1,\; a_2,\; a_3$ are these three matrices, then we get that
$(1 - \frac{1}{p^3})n^2 + \frac{m}{p^3} - O(n^{\frac{7}{4}})$ linear forms
vanish on $I,\; (a_1-a_3),\; (a_2 - a_3)$, 
and that $[(a_1-a_3),(a_2 - a_3)]$ is an invertible matrix, which is what we
wanted to prove.

So assume w.l.o.g. that the linear forms
$\mu_{n^2 +1} ,\; \ldots,\; \mu_{n^2 + r} \;,$ vanish on $I$
(beside those among $\mu_{1} ,\; \ldots,\; \mu_{n^2}$ that vanish on it) 
where $r\geq \frac{m- n^2}{p} - O(n^{\frac{5}{3}})$.
Let $a_1,\; \dots,\; a_k$ be the matrices guaranteed by 
lemma~\ref{k-commutator}. Consider the following vectors:
$\forall1 \leq i \leq k \;,\; $
$$v_i = (\mu_{n^2 +1}(a_i) ,\; \ldots,\; 
\mu_{n^2 + r}(a_i)) \in {\F(p)}^r \;.$$
According to lemma~\ref{intersect} three of these vectors,
namely $v_1,\; v_2,\; v_3$,  agree on at least
$\frac{r}{p^2}- \frac{3r}{pk} $ coordinates. 
Therefore there are $\frac{r}{p^2}- \frac{3r}{pk} $ linear forms among
$\mu_{n^2 +1},\; \ldots,\; \mu_{n^2 + r}$ that vanish on 
$a_1-a_3$ and $a_2 - a_3$. In addition there are
$n^2 - 2{k \choose 3}n$ linear forms among
$\mu_1,\ldots,\mu_{n^2}$ that vanish on $a_1,a_2,a_3$, 
hence there are $n^2 - 2{k \choose 3}n$ linear forms among
$\mu_1,\; \ldots,\; \mu_{n^2}$ that vanish on $a_1-a_3$ and on $a_2-a_3$.
Let $a=a_1-a_3,\; b=a_2-a_3$.

We get that there are 
$\frac{r}{p^2} - \frac{3r}{pk}$ linear forms among
$\mu_{n^2 +1},\; \ldots,\; \mu_{n^2 + r}$ that vanish on $I,\; a,\; b$.
Since $n^2 - O(n^{\frac{5}{3}})$ of the first $n^2$ $\mu_i$'s vanish on $I$,
we get that at least $n^2 - 2{k \choose 3}n - O(n^{\frac{5}{3}})$ 
of the first $n^2$ $\mu_i$'s vanish on $I,\; a, \; b$.
Putting it all together we get that at least
$$n^2 - 2{k \choose 3}n - O(n^{\frac{5}{3}}) + \frac{r}{p^2} -
\frac{3r}{pk}$$
linear forms among $\mu_1,\; \dots,\; \mu_m$ vanish on $I,\; a,\; b$. 
Since $r= \frac{m-n^2}{p} - O(n^{\frac{5}{3}})$, and
$k=n^{\frac{1}{4}}$,  we get that at least
$$(1-\frac{1}{p^3})n^2 + \frac{m}{p^3} - O(n^{\frac{7}{4}}) $$
of the $\mu_i$'s vanish on them. This concludes the proof of 
lemma~\ref{2-commutator}.
\end{proof}

Putting everything together we get by lemma~\ref{bla} and 
lemma~\ref{2-commutator} that:
$$m \geq 1.5 n^2 + (1-\frac{1}{p^3})n^2 + \frac{m}{p^3} -
O(n^{\frac{7}{4}}) \;.$$
Therefore
$$ m \geq (2.5 + \frac{1.5}{p^3 -1})n^2 - O(n^{\frac{7}{4}}) \;. $$
This concludes the proof of theorem~\ref{main_finite}.
\end{proofof}

\section{Acknowledgment}
I would like to thank Michael Ben-Or and Avi Wigderson 
for helpful conversations. 
I would also like to thank Ran Raz for commenting on an 
earlier draft of the paper (in his words: Would you like to hear my comments
or do you just want to rewrite everything).





\end{document}